\documentclass[aps,prl,superscriptaddress,twocolumn,floatfix,a4paper]{revtex4}

\usepackage{graphicx,graphics,epsfig}   
\usepackage{dcolumn}   
\usepackage{amsmath} 
\usepackage{verbatim}   
\usepackage{color}       
\usepackage{subfigure}  
\usepackage{times,natbib}
\usepackage{amsmath,amsfonts,amssymb,graphics,graphicx,epsfig,color,times,natbib}

\newcommand{\be}{\begin{equation}}
\newcommand{\ee}{\end{equation}}
\newcommand{\ba}{\begin{eqnarray}}
\newcommand{\ea}{\end{eqnarray}}
\newcommand{\ban}{\begin{eqnarray*}}
\newcommand{\ean}{\end{eqnarray*}}

\begin{document}


\title{Closed sets of non-local correlations}

\author{Jonathan Allcock}
\affiliation{Department of Mathematics, University of Bristol, Bristol, BS8 1TW, United Kingdom}
\author{Nicolas Brunner}
\affiliation{H.H. Wills Physics Laboratory, University of Bristol, Tyndall Avenue, Bristol, BS8 1TL, United Kingdom}
\author{Noah Linden}
\affiliation{Department of Mathematics, University of Bristol, Bristol, BS8 1TW, United Kingdom}
\author{Sandu Popescu}
\affiliation{H.H. Wills Physics Laboratory, University of Bristol, Tyndall Avenue, Bristol, BS8 1TL, United Kingdom}
\author{Paul Skrzypczyk}
\affiliation{H.H. Wills Physics Laboratory, University of Bristol, Tyndall Avenue, Bristol, BS8 1TL, United Kingdom}
\author{Tam\'as V\'ertesi}
\affiliation{ Institute of Nuclear Research of the Hungarian Academy of Sciences
H-4001 Debrecen, P.O. Box 51, Hungary}

\date{Wednesday 26 August 2009}


\begin{abstract}

We introduce a fundamental concept -- closed sets of correlations -- for studying non-local correlations. We
argue that sets of correlations corresponding to information-theoretic principles, or more generally to
consistent physical theories, must be closed under a natural set of operations. Hence, studying the closure of
sets of correlations gives insight into which information-theoretic principles are genuinely different, and
which are ultimately equivalent. This concept also has implications for understanding why quantum non-locality
is limited, and for finding constraints on physical theories beyond quantum mechanics.

\end{abstract}

\maketitle

Correlations are a central concept in physics. While in classical physics correlations must satisfy two fundamental principles - causality and locality - in quantum mechanics (QM) the latter must be abandoned. This remarkable feature, known as quantum non-locality, is at the heart of quantum information processing and allows tasks to be performed which would be impossible classically, such as secure cryptography \cite{ekert} and the reduction of communication complexity \cite{redComm}.

However, non-local correlations stronger than those allowed by QM can also respect relativistic causality \cite{PR}. These non-signaling post-quantum correlations have been subject to intensive research \cite{barrett,NS,NS_crypto,CC,noah,emergence,distillation,IC}, and were shown to have strong information-theoretic consequences, allowing for powerful tasks - impossible in QM - to be peformed. For instance, certain post-quantum correlations collapse communication complexity \cite{CC,distillation}; allow for better-than-classical `non-local computation' \cite{noah}; and violate `information causality' \cite{IC}.

Here we introduce a fundamental concept -- closed sets of correlations -- underlying the structure of non-local correlations. We argue that physically significant sets of correlations must be closed under a natural class of operations.

The immediate relevance of this concept if two-fold. First, we note that all information-theoretic principles
correspond to closed sets of correlations. For instance, the set of correlations that do not make communication
complexity trivial is closed. If two different information-theoretic principles turn out to correspond to the
same closed set then they are in fact equivalent as far as the resources needed to implement them are concerned.
Therefore, studying the closure of sets of correlations gives insight into which information-theoretic
principles are genuinely different, and which are ultimately equivalent. This also leads one to ask: is there an
infinite number of closed sets or only finitely many? If it was found that only a small number of closed sets
can exist, then most information-theoretic principles would turn out to be the same.

Even more importantly, correlations allowed by any self-consistent physical theory must form a closed set. For instance in classical mechanics, it is impossible to generate non-local correlations from local ones. Similarly, post-quantum correlations cannot be generated within the framework of QM. From this perspective, the concept of closure gives insight into why quantum non-locality is limited, and provides a platform for finding novel physical theories beyond QM.

We will work in the formalism of non-signaling boxes \cite{barrett}. The natural set of operations we consider correspond to \emph{wirings} \cite{short,john}, which can be thought of as classical circuitry used to locally connect several non-signaling boxes in order to obtain a new box. A set of boxes $\mathcal{R}$ is said to be \emph{closed under wirings} when all boxes obtainable by wiring boxes in $\mathcal{R}$ are also contained in $\mathcal{R}$. Interestingly, we shall see that finding closed sets is a non-trivial task.

\textbf{\emph{Preliminaries.}} Let us recall that bipartite non-signaling correlations can be conveniently viewed in terms of black boxes shared between two parties, Alice and Bob. Alice and Bob input variables $x$ and $y$ at their ends of the box respectively, and get outputs $a$ and $b$.  The behavior of a given box is fully described by a set of joint probabilities $P(ab|xy)$. We focus on the case of binary inputs and outputs, i.e. $a,b,x,y\in\left\{0,1\right\}$. In this case, the full set $\mathcal{NS}$ of non-signalling boxes  forms an 8-dimensional polytope \cite{barrett} which has 24 vertices: 8 extremal non-local boxes and 16 local deterministic boxes. The extremal non-local correlations have the form:
\begin{equation}
P_{\text{NL}}^{\mu\nu\sigma}(ab|xy) =
    \begin{cases}
        \frac{1}{2} & \text{if} \quad \text{$a \oplus b = xy\oplus\mu
x\oplus\nu y\oplus\sigma $} \\
        0 & \text{otherwise}
    \end{cases}
\end{equation}
where $\mu,\nu,\sigma\in\left\{0,1\right\}$, and the canonical Popescu-Rohrlich (PR) \cite{PR,barrett} box corresponds to $\text{PR}=
P_\text{NL}^{000}$. Similarly, the local deterministic boxes are given by
\begin{equation}
P_{\text{L}}^{\mu\nu\sigma\tau}(ab|xy) =
    \begin{cases}
        1 & \text{if} \quad\text{$a = \mu x \oplus\nu$} \quad \text{$b =
\sigma y\oplus\tau$} \\
        0 & \text{otherwise}
    \end{cases}
\end{equation}
The set $\mathcal{L}$ of local boxes forms a subpolytope of the full non-signalling polytope. $\mathcal{NS}$ has 16 facets (positivity facets), and $\mathcal{L}$ has 8 additional facets, which correspond to the 8 symmetries of the Clauser-Horne-Shimony-Holt (CHSH) Bell inequality \cite{chsh}: $\text{CHSH} \equiv E_{00}+E_{01}+E_{10}-E_{11}\leq 2$, where $E_{xy} \equiv P(a=b|xy)-P(a\neq b|xy)$ are the correlators. The set of quantum boxes $\mathcal{Q}$, i.e.\ correlations obtainable by performing local measurements on a quantum state satisfies $\mathcal{L} \subset\mathcal{Q}\subset\mathcal{NS}$. Quantum non-locality is limited by Tsirelson's bound \cite{tsirelson2}, given by $\text{CHSH}\leq B_Q = 2\sqrt{2}$. $\mathcal{Q}$ is a convex body, though not a polytope; it has a curved boundary for which no closed form is known \cite{miguel}.

\begin{figure}[t]
  \includegraphics[width=0.6\columnwidth]{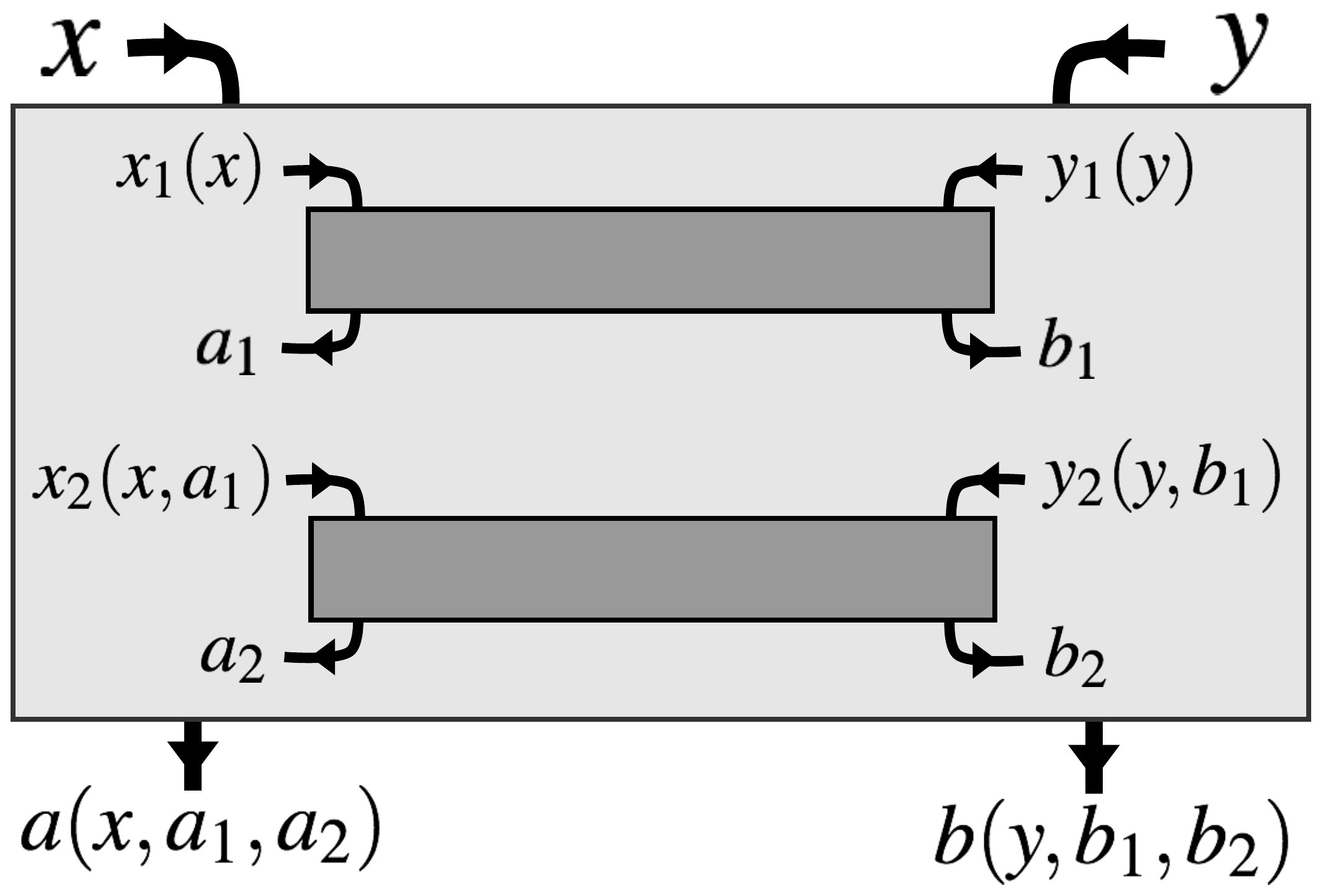}
  \caption{Wiring several boxes together to produce a new box.}
\label{fig:wirings}
\end{figure}

\textbf{\emph{Wirings.}}
Suppose that Alice and Bob share $N$ non-signaling boxes.
Since each box has binary inputs and outputs,
Alice and Bob can `wire' the boxes together using classical circuitry to produce a new binary-input/binary-output box (see Fig.~\ref{fig:wirings}). The inputs and outputs of the $j$th box are denoted $x_j,y_j,a_j,b_j$. Since the inputs of the $j$th box can depend on the outputs of boxes $1,...,j-1$, a wiring is fully determined by specifying Boolean functions for the inputs to each box, $x_j(x,a_1,...,a_{j-1})$ and $y_k(y,b_1,...,b_{k-1})$, as well as Boolean functions for the final output bits, $a(x,a_1,...,a_N)$ and $b(y,b_1,...,b_N)$. Since the boxes are non-signaling, when Alice inputs a bit in a given box, she gets an output immediately, even if Bob is yet to input a bit in his end of the box. This allows for interesting situations, in which Alice's and Bob's chronological orderings of their boxes are different.

\emph{Distillation Wirings}. Starting from several copies of a non-local box with a given CHSH value, it is possible - via wirings - to obtain a final box which has a larger CHSH value.  This is known as non-locality distillation, recently discovered in \cite{foster}, and improved in \cite{distillation}. Here we present a novel two-box distillation protocol. Alice proceeds as follows: $x_1=x$, $x_2=x\oplus a_1\oplus 1$, $a=a_1\oplus a_2\oplus 1$; and Bob: $y_1=y$, $y_2=y b_1$, $b=b_1\oplus b_2 \oplus 1$. This protocol can distill efficiently a class of boxes we term `correlated non-local boxes':
\begin{equation}\label{correlated}
    P_{\text{NL}}^{\text{c}}(\epsilon) = \epsilon \text{PR} + (1-\epsilon) P_L^{0101},
\end{equation}
where $0<\epsilon \leq 1$.  These boxes have CHSH value $2(1+\epsilon)$. By applying the above protocol to two copies of a box $P_{\text{NL}}^{\text{c}}(\epsilon)$, one obtains a box $P_{\text{NL}}^{\text{c}}(\epsilon')$, with $\epsilon'= 2\epsilon-\epsilon^2$.  Since $\epsilon'>\epsilon$, the protocol distills non-locality (as measured by the CHSH value). In the asymptotic limit, all boxes \eqref{correlated} are distilled to the maximally non-local PR box.

\emph{AND Wirings}. Another interesting class of wirings involves Alice and Bob inputting $x$ and $y$ respectively into each of their $N$ boxes, and computing the logical AND of their outputs; i.e. $x_j=x$ for $j\in\{1,...,N\}$ and $a=\prod_{j=1}^N a_j$; similarly for Bob. If Alice and Bob share $N$ copies of an initial box $P(ab|xy)$, the final box $P^{\prime}(ab|xy)$ obtained from such a wiring is easily characterized \footnote{See supplementary material.}. AND wirings can be used for distillation, but will primarily be useful for showing that certain sets of correlations are not closed.

\begin{figure}[t]
  \includegraphics[width=\columnwidth]{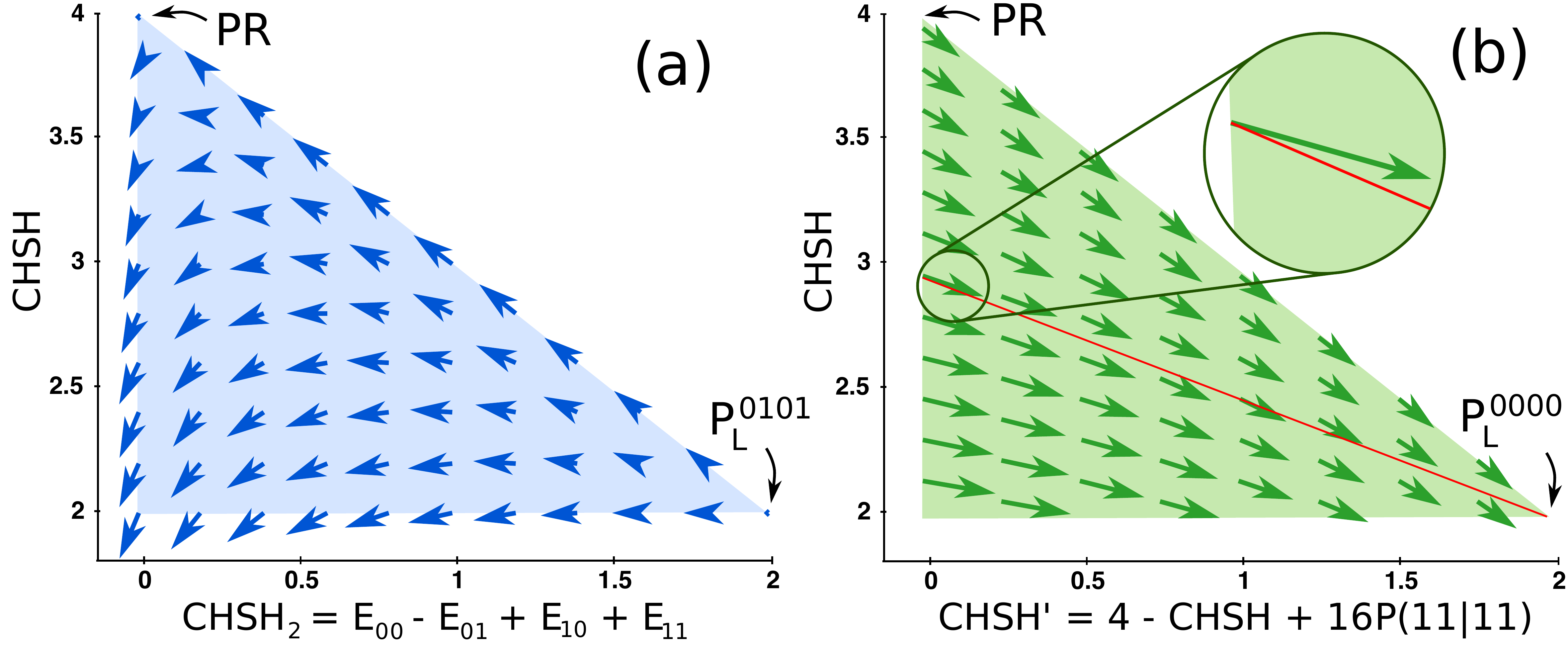}
  \caption{Vector fields for (a) the distillation wiring, (b) the (2-box) AND wiring, in certain 2D sections of the non-signaling polytope. Clearly, correlated non-local boxes \eqref{correlated} are distilled in (a). The AND wiring (b) maps isotropic boxes \eqref{Noisy2} above their `chord' (red line); thus restricted polytopes in case study b) (see text) are not closed. This effect can only be seen in specific projections of the polytope.}
\label{fig:vector}
\end{figure}

\emph{Discrete maps.} When studying the closure of sets of correlations, it is essential to understand how boxes can be `moved around' in the non-signaling polytope using wirings. A useful approach \cite{distillation} is to look at discrete maps $\mathcal{T}$ which take multiple copies of an initial box $B_i$ - via wirings - to a final box $B_f$, i.e. $\mathcal{T}(B_i)=B_f$. Then, all standard techniques for studying discrete maps can be used. The asymptotic behavior is characterized by the fixed points of the map, the stability of which can be checked by computing the eigenvalues of the Jacobian. Moreover, plotting the map's vector field provides some intuition about the action of a wiring protocol in a given section of the non-signaling polytope. Fig. \ref{fig:vector} shows the vector fields for both wirings (distillation, AND) described above.

\textbf{\emph{Closure under wirings.}} Consider a theory where Alice and Bob have access to boxes from some set $\mathcal{R}$. Given that, by wiring multiple boxes together, they can produce a new non-signaling box, it is natural to ask whether the resultant box is also in $\mathcal{R}$. We will call a set of correlations $\mathcal{R}$ \emph{closed under wirings} if it is impossible to generate, by wiring together boxes contained in $\mathcal{R}$, a box $B$ that is not contained in $R$. We define the  \emph{closure of $\mathcal{R}$ under wirings} to be the smallest closed set $\mathcal{C}$ such that $\mathcal{R} \subseteq \mathcal{C}$. The sets $\mathcal{L}$, $\mathcal{Q}$ and $\mathcal{NS}$ are all examples of closed sets. Note that $\mathcal{L}$ is the smallest possible set of correlations closed under wirings; indeed all deterministic boxes can always be generated using a trivial (deterministic) wiring, which implies that any closed set must include $\mathcal{L}$.

Studying the relation between different closed sets leads to further interesting concepts, such as an \emph{irreversibility} in the flow of boxes. Consider two closed sets $\mathcal{C}$ and $\mathcal{C}^{\prime}$ such that $\mathcal{C} \subset \mathcal{C}^{\prime}$. Then the set of boxes $\mathcal{\widetilde{R}} \equiv \mathcal{C}^{\prime}\backslash\mathcal{C}$ (i.e. boxes in $\mathcal{C}^{\prime}$ but not in $\mathcal{C}$) forms an \emph{island}, in the sense that when a box in $\mathcal{\widetilde{R}}$ is mapped out of $\mathcal{\widetilde{R}}$, it can never be mapped back into $\mathcal{\widetilde{R}}$ again. Thus the boundary between $\mathcal{C}$ and $\mathcal{\widetilde{R}}$ acts like a horizon, restricting the flow of boxes. The set $\mathcal{NS}\backslash\mathcal{Q}$ is an example of an island.

\emph{Case studies.} It is interesting to ask whether there exist other sets which are closed under wirings and, if so, what their structure is. Given that both $\mathcal{L}$ (no non-locality) and $\mathcal{NS}$ (maximal non-locality) form closed \emph{polytopes}, it is tempting to look for closed polytopes which have limited non-locality.  We now attempt to construct such a polytope, and show that the most natural candidates fail. Then, we move to convex sets which are not polytopes; in particular we consider the sets of Uffink \cite{Uffink02} and Pitowsky \cite{Pito08}. Again, we show that neither of these two sets are closed under wirings.

\begin{figure}[t]
  \includegraphics[width=\columnwidth]{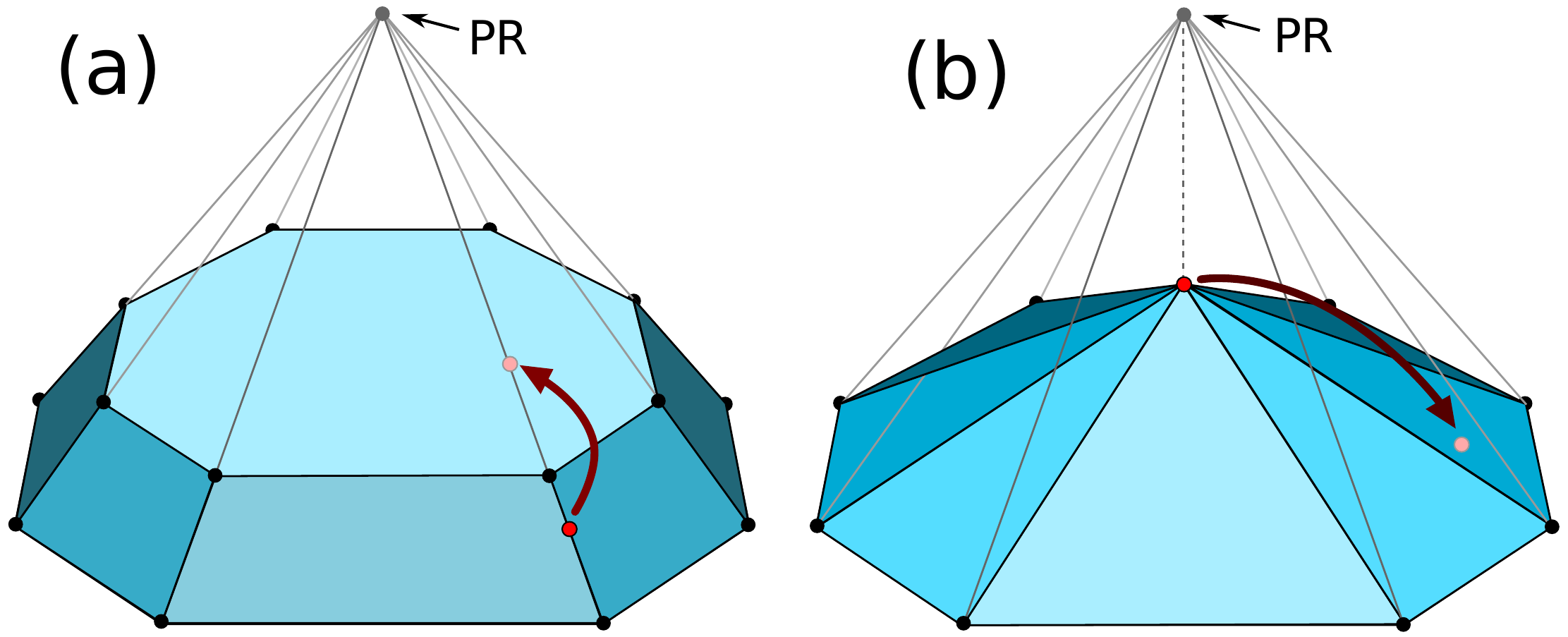}
  \caption{Theories with limited non-locality: (a) limiting the CHSH value by a cut-off; (b) making the extremal PR boxes noisy. Neither set is closed since - as the arrows indicate - boxes which lie inside each set can be mapped to boxes which lie outside. Note that these pictures are illustrative; the actual polytopes are 8-dimensional.}
\label{fig:polytope_VII}
\end{figure}

\emph{a. Limiting the CHSH value.} The simplest way of constructing a non-signaling set with limited non-locality is to bound the CHSH value. That is, we consider $\mathcal{NS}$ and then discard all boxes with a CHSH value larger than some cut-off $S$, where $2<S<4$ (see Fig.~\ref{fig:polytope_VII}(a)).  Formally, the set of boxes in such a theory forms a restricted polytope $\mathcal{R}_a^S$; its facets are the 16 positivity facets plus 8 CHSH-like facets of the form $\text{CHSH} \leq S$. Its vertices are the 16 local deterministic boxes plus 64 non-local vertices given by \ba\label{Noisy1} P_{\text{NL}}^{\mu\nu\sigma,\alpha\beta\gamma}(\epsilon) = \epsilon P_\text{NL}^{\mu\nu\sigma} +  (1-\epsilon)P_{\text{L}}^{\alpha\beta\gamma\delta}, \ea with $\delta = (\alpha\oplus \mu)(\gamma \oplus \nu)\oplus \beta \oplus \sigma$; the indices $\mu,\nu,\sigma$ run over all symmetries of the PR box, and the indices $\alpha,\beta,\gamma$ run over the 8 deterministic boxes sitting on the CHSH facet below each PR box. Note that $\epsilon = \frac{S}{2}-1$.

However, $\mathcal{R}_a^S$ is not closed under wirings for any value of $2<S<4$, since any box lying on a one-dimensional edge of $\mathcal{NS}\backslash\mathcal{L}$ (of the form \eqref{Noisy1}) can be distilled arbitrarily close to a PR box using our distillation protocol; note that for each edge a suitable symmetry of the protocol must be used. Thus, the closure of $\mathcal{R}_a^S$ is $\mathcal{NS}$. More generally this implies that a set of boxes - if it is to be closed under wirings - cannot contain any box lying on a one-dimensional edge of $\mathcal{NS}\backslash\mathcal{L}$.

\emph{b. Making the PR boxes noisy.} Given that $\mathcal{NS}$ consists of local and non-local vertices, another natural way of defining a polytope with limited non-locality is to keep all 16 local vertices, and modify the nonlocal vertices by adding isotropic (white) noise (see Fig.~\ref{fig:polytope_VII}(b)). In other words, the extremal nonlocal vertices of such a polytope have the form
\begin{equation} \label{Noisy2}
     \text{P}_{\text{NL}}^{\mu\nu\sigma}(\epsilon) = \epsilon \text{P}_\text{NL}^{\mu\nu\sigma} +(1-\epsilon) \openone ,
\end{equation}
where $\openone$ is the maximally mixed box. We denote this restricted polytope $\mathcal{R}_b^S$. In this case non-locality is limited by $S = 4\epsilon$. Note that $\mathcal{R}_b^S$ does not contain any non-local boxes lying on a one-dimensional edge of $\mathcal{NS}\backslash\mathcal{L}$.

It is clear that such a polytope can (potentially) be closed only if isotropic boxes cannot be distilled. So far no distillation protocol has been found for isotropic boxes. For quantum realizable isotropic boxes (with $4\epsilon\leq B_Q$) severe restrictions have been proven in \cite{dejan}, while \cite{toni_pur} proved that there is no two-copy distillation protocol for isotropic boxes.

However, it can be shown that the sets $\mathcal{R}_b^S$ are not closed using an alternative method. Using the software LRS \footnote{D. Avis, http://cgm.cs.mcgill.ca/~avis/C/lrs.html} we found all the facets of $\mathcal{R}_b^S$, of which there are 80: 64 new facets in addition to the original 16 positivity facets. One particular new facet is given by
\begin{equation}\label{Facet}
     I(q) \equiv \text{CH}+q P(11|11) \geq 0,
\end{equation}
where $\text{CH}=1- P(11|00)-P(00|10)-P(00|01)+P(00|11)$ is the Clauser-Horne expression (equivalent to CHSH), and $q=\frac{2(2\epsilon-1)}{1-\epsilon}$. Note that \eqref{Facet} is a `tilted' CH inequality; $I(q=0)=\text{CH}$, while $I(q\rightarrow \infty)$ is a positivity facet.

It turns out that one can generate a box violating inequality \eqref{Facet} by applying the AND wiring to $N$ copies of an isotropic box $\text{P}_{\text{NL}}^{000}(\epsilon)$. The final box (which does not have a greater CHSH value than the original boxes) is found to violate \eqref{Facet} whenever $2^{1-N}-3z_+^N + (1+q)z_-^N<0$, where $z_{\pm}=\frac{1\pm \epsilon}{4}$. For $N=2$ (best case), \eqref{Facet} is satisfied for $\frac{2}{3}<\epsilon <1$. Thus all restricted polytopes $\mathcal{R}_b^S$ for which $\frac{8}{3}<S<4$ are not closed; moreover a two-box AND wiring is sufficient to generate a box lying outside the original set. We exhaustively checked that $\mathcal{R}_b^S$ is closed under all two-box wirings for $S<\frac{8}{3}< B_Q$. However, we conjecture that these sets are not closed under more general wirings, but were
not able to prove it.

Coming back to the sets $\mathcal{R}_b^S$ with $\frac{8}{3}<S<4$, we have shown explicitly how - via a two-box AND wiring - to generate a particular box that lies outside $\mathcal{R}_b^S$. By symmetry, it is possible to generate 64 new boxes which each violate one of the new facets. We can now consider a new polytope, $\mathcal{R}_b^{S(2)}$, obtained by taking the convex hull of $\mathcal{R}_b^S$ and the 64 newly generated boxes. It is natural to ask whether this larger set $\mathcal{R}_b^{S(2)}$ is closed under wirings or not. If not, one can again form a new polytope $\mathcal{R}_b^{S(3)}$ by adding the newly generated vertices and so on.  Even under this very restricted class of wirings (AND wirings applied to $N$ isotropic boxes) we find that this procedure can be iterated multiple times (the number of times increases with $S$), which leads us to conjecture that the closure of $\mathcal{R}_b^S$ has a boundary with curved sections [24].

\emph{c. Uffink and Pitowsky sets.} In studying quantum non-locality, different subsets of $\mathcal{NS}$ have been introduced, such as Uffink's set \cite{Uffink02}, characterized by the quadratic form:

\ba\label{Uffink} (E_{00}+E_{10})^2 + (E_{01}-E_{11})^2 \leq 4. \ea Using the distillation protocol presented above, it can be shown that Uffink's set is not closed [24]; this analysis also applies to the convex set of Pitowsky \cite{Pito08}. Notably, Uffink's set emerges \cite{IC_Qboundary} from the principle of information causality (IC) \cite{IC}; that is, any box violating inequality \eqref{Uffink} also violates IC. Our result implies that the set of correlations satisfying IC can be at most the largest closed subset of Uffink's set. Furthermore, we see that neither set (Uffink or Pitowsky) can correspond to any information-theoretic principle or task.

\textbf{\emph{Discussion.}} We introduced the concept of closed sets of correlations. We argued that studying these sets is of fundamental importance, since all information-theoretic principles, as well as all self-consistent physical theories \footnote{Note that the model of \cite{emergence} is conceptually different; the set of \emph{genuine boxes} does not need to be closed under wirings.}, correspond to closed sets of correlations.

We investigated closure under a natural class of operations called wirings, and showed that identifying closed
sets is a highly non-trivial problem. Our results also illustrate the relevance of closure to information
theoretic tasks. By showing that the convex set of Uffink is not closed, we could find novel constraints on the
set of correlations satisfying the principle of information causality.

Moreover these ideas provide new insight into the origin of the boundary between quantum and post-quantum
correlations. For instance, if QM was the only closed set (other than the full set of non-local correlations) containing non-local correlations, then this would be enough to single out the quantum set
$\mathcal{Q}$. However, this turns out not to be the case. Recently, Navascues and Wunderlich
\cite{macro_locality} have shown that the set $\mathcal{Q}^1$ \cite{miguel} (an approximation to $\mathcal{Q}$)
is also closed under wirings. Non-locality in $\mathcal{Q}^1$ is limited by Tsirelson's bound, but
$\mathcal{Q}^1$ strictly contains $\mathcal{Q}$. Nevertheless it could still be that $\mathcal{Q}$ is the
smallest closed set with non-locality limited by Tsirelson's bound. Going one step further, $\mathcal{Q}$ could
in fact be the smallest possible closed set featuring non-locality. To test these ideas it would be interesting
to see if there exist closed sets for which non-locality is limited by a different value than Tsirelson's bound;
so far, $\mathcal{Q}$ and $\mathcal{Q}^1$ are the only sets with limited non-locality known to be closed under
wirings.

Another interesting issue is understanding the structure of closed sets. Our findings lead us to conjecture that all closed sets of correlations with limited non-locality have a curved boundary. If true, this would imply that the local set and the non-signaling set are the only closed sets that form a polytope.

We also note that wirings are a particular subclass of the most general operations that can be performed on non-signaling boxes. Closure under more general operations, such as \emph{couplers} \cite{emergence,couplers} the analogue of quantum joint measurements, may give further restrictions on the class of closed sets.

Finally, from a much more general perspective, the concept of closure may also give us a glimpse of what lies beyond QM. Indeed it is plausible that in the future QM will be superseded by a more general theory. Though defining explicitly such a theory is highly challenging, consistency requires this theory to correspond to a closed set of correlations.

\emph{Acknowledgements.} We acknowledge support from the Dorothy Hodgkins Foundation (JA), the Swiss National Science Foundation (NB), the UK EPSRC project `QIP IRC' and the EU project QAP (NB, NL, SP and PS), and the J\'anos Bolyai Programme of the Hungarian Academy of Sciences (TV).

\bibliographystyle{prsty}
\bibliography{C:/BIB/thesis}

\section{Appendix A}
\textit{AND wirings applied to $N$ boxes}. If Alice and Bob share $N$ copies of an initial box $P(ab|xy)$, the final box $P^{\prime}(ab|xy)$ obtained from applying the AND wiring is given by
\begin{align*}
P^{\prime}(11|xy) &= P\left(11|xy\right)^N, \\
P^{\prime}(01|xy) &= \left[P\left(01|xy\right) + P\left(11|xy\right)\right]^N - P\left(11|xy\right)^N,\\
P^{\prime}(10|xy) &= \left[P\left(10|xy\right) + P\left(11|xy\right)\right]^N - P\left(11|xy\right)^N, \\
P^{\prime}(00|xy) &= 1 - P^{\prime}\left(01|xy\right) - P^{\prime}\left(10|xy\right)- P^{\prime}\left(11|xy\right).
\end{align*}

\begin{figure}[h!]
  \includegraphics[width=\columnwidth]{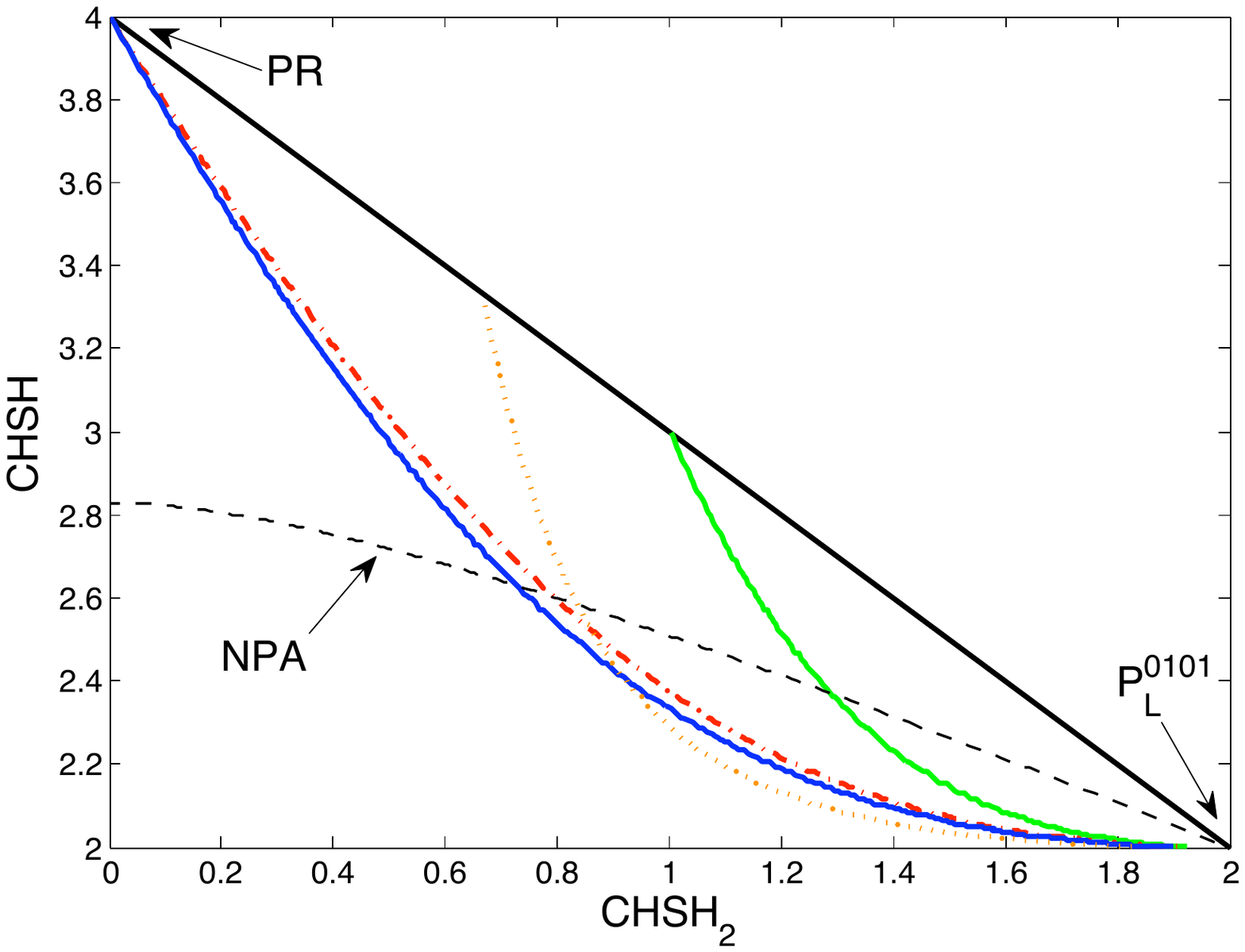}\\
  \caption{Comparison of distillation protocols in the 2D slice of the polytope defined by \eqref{PRP1Id1}. The different lines delimit the set of boxes that can be distilled in one (or more) iterations of a given protocol. The solid blue line corresponds to the distillation protocol presented in the main text; all boxes above this line can be distilled using this distillation protocol. The dash-dotted red line is for the protocol of Ref. \cite{distillation}. The dotted orange line is for the two-box AND protocol. The green solid line corresponds to the protocol of Ref. \cite{foster}. The black dashed line is an upper bound on the set of quantum correlations, derived by Navascues-Pironio-Acin (NPA) \cite{miguel}.}
\label{fig:Comp_distill}
\end{figure}

As mentioned in the main text, the AND protocol can also be used to distill non-locality. Fig. \eqref{fig:Comp_distill} shows a comparison of all known two-box distillation protocols. We consider a section of the non-signaling polytope, given by boxes of the form:
\begin{equation}\label{PRP1Id1}
	P_{\text{NL}}^{\text{c}}(\epsilon,\gamma) = \epsilon \text{PR} + \gamma  P_L^{0101} +  (1-\epsilon-\gamma)\openone
\end{equation}

As an aside, note that the distillation protocol that we presented above slightly improves on the protocol of \cite{distillation} for noisy correlated non-local boxes, i.e. for boxes $P_{\text{NL}}^{\text{c}}(\epsilon,\gamma)$ with $\epsilon +\gamma<1$. For correlated non-local boxes ($\epsilon +\gamma=1$), i.e. on the edge of the polytope, both protocols perform equally.

\begin{figure}[h!]
  \includegraphics[width=\columnwidth]{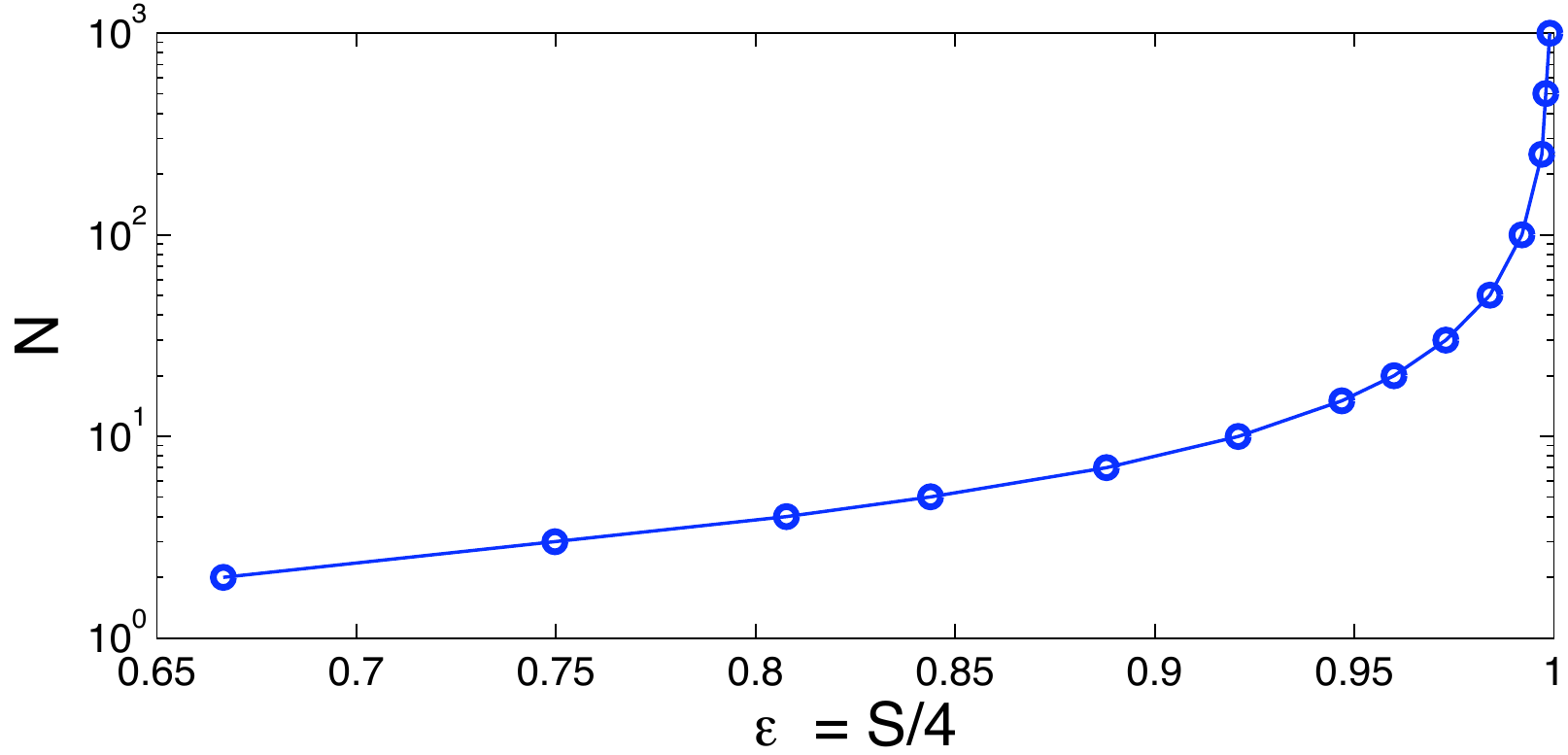}\\
  \caption{Number of boxes $N$ that can be generated outside the restricted polytope $\mathcal{R}_b^S$. Each new box is obtained by applying the AND wiring to $N$ copies of the isotropic box \eqref{Noisy2}, characterized by $\epsilon$.}
\label{fig:NandOut}
\end{figure}

\textit{AND protocol and case study b}. Let us consider again the closure of the restricted polytope $\mathcal{R}_b^S$. Starting from $N$ copies of an isotropic box \eqref{Noisy2} characterized by $\epsilon$, it can be checked that the final box violates the facet $I(q)$ of $\mathcal{R}_b^S$ whenever
\begin{equation}
	\frac{1}{2^{N-1}}-3 \left( \frac{1+\epsilon}{4}\right)^N + q \left( \frac{1-\epsilon}{4}\right)^N <0.
\end{equation}
This leads to a relation between the number of new boxes $N$ that can be generated outside $\mathcal{R}_b^S$ and the CHSH value of the initial isotropic boxes, i.e. $S=4\epsilon$. As can be seen from Fig. \eqref{fig:NandOut}, the number of boxes that can be generated outside $\mathcal{R}_b^S$ rapidly increases as $\epsilon \rightarrow 1$.

Thus, for each $N$, we obtain a polytope $\mathcal{R}_b^{S(N)}$, which is the convex hull of $\mathcal{R}_b^S$ and all the newly generated boxes lying outside $\mathcal{R}_b^S$. Interestingly, it appears that $\mathcal{R}_b^{S(N-1)}\subset \mathcal{R}_b^{S(N)}$; that is, for each $N$, the box generated (and its symmetries) lie actually outside the polytope obtained in the $N-1$ case; we checked this numerically for $3\leq N\leq 10$ and $\epsilon=0.95$. Thus it appears that when $\epsilon \rightarrow 1$ we can generate a polytope with an arbitrarily large number of extremal points. Note that it is only in the limit $\epsilon \rightarrow 1$ that the boundary becomes a smooth curve.

\section{Appendix B}
\emph{Distilling out of the Uffink set.} Here we show that Uffink's set in not closed under wirings, using the distillation protocol introduced in the main text.

We consider a section of the non-signaling polytope, given by boxes of the form:
\begin{equation}\label{PRP1Id2}
	P_{\text{NL}}^{\text{c}}(\epsilon,\gamma) = \epsilon \text{PR} + \gamma  P_L^{0101} +  (1-\epsilon-\gamma)\openone
\end{equation}
It will be convenient to characterize these boxes by their four correlators $E_{xy} \equiv P(a=b|x,y)-P(a\neq b|x,y)$; here we have $E_{00}=E_{01}=E_{10}=\epsilon+\gamma$ and $E_{11}=\gamma-\epsilon$. After applying the distillation protocol to two copies of the box \eqref{PRP1Id2}, we obtain a final box given by
\ba\nonumber B_f &=& \frac{\epsilon}{4}(3\epsilon +7\gamma+1)\text{PR}  + \frac{\epsilon}{4}(1-\epsilon-\gamma) P_{\text{NL}}^{011} \\ & & + \gamma^2 P_L^{0101} + (1-\epsilon-\gamma)(1+\frac{\epsilon}{2}+\gamma) \openone \ea The correlators of $B_f$ are
\ba\nonumber E_{00}^f&=&E_{10}^f= (\epsilon+\gamma)^2 \\ E_{01}^f &=& \frac{1}{2}\left[ (\epsilon+\gamma)^2 + \epsilon \gamma +\gamma^2 + \epsilon \right] \\\nonumber E_{11}^f &=& -\frac{1}{2} \left[ (\epsilon+\gamma)^2 + \epsilon \gamma -3\gamma^2 + \epsilon \right]\ea
Now, we impose that $B_f$ must lie outside Uffink, i.e.

\begin{figure}[h!]
  \includegraphics[width=\columnwidth]{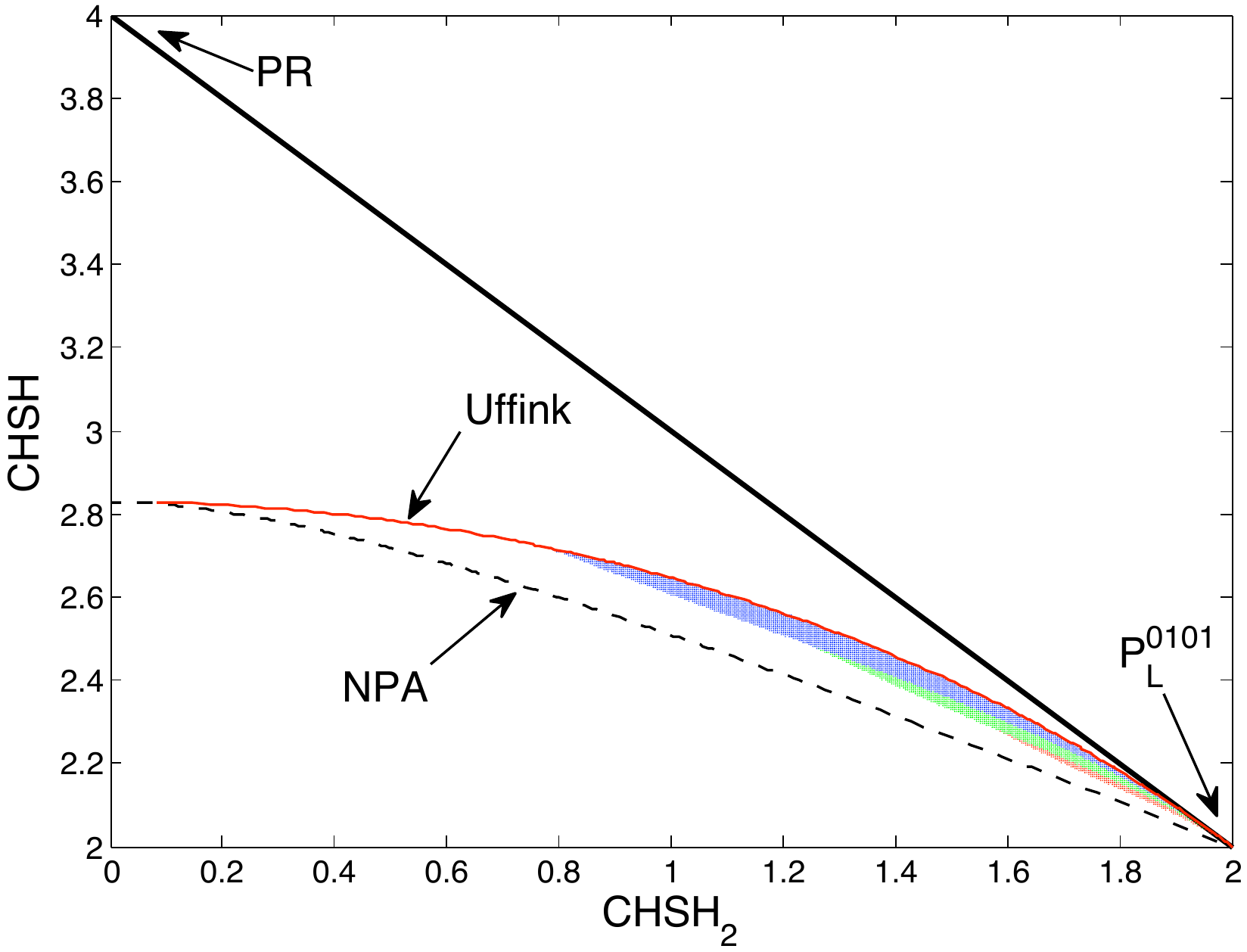}\\
  \caption{2D section of the non-signaling polytope. The solid red line is the boundary of Uffink's set, and the dashed black line is the upper bound on the quantum set derived by NPA \cite{miguel}. The shaded region are boxes, initially inside Uffink's set, that are mapped outside using the distillation protocol; more specifically, in one iteration of the protocol (blue region), two iterations (green region), and three iterations (red region). The lower boundary of the blue region (one iteration) is given by equation \eqref{UffDist}.}
\label{fig:Uffink}
\end{figure}

\ba\label{Uffink2} (E_{00}^f+E_{10}^f)^2 + (E_{01}^f-E_{11}^f)^2 > 4 \ea Next, using the relation $\epsilon = \frac{E_{00}-E_{11}}{2}$ and $\gamma = \frac{E_{00}+E_{11}}{2}$, we can rewrite \eqref{Uffink2} in terms of the correlators of the initial box and obtain

\ba\label{UffDist} 4 E_{00}^4 + \left[ E_{00}^2 +\frac{1}{2}(E_{00} - E_{11} - E_{11}^2-E_{00}E_{11}) \right]^2 >4 \ea It turns out that the previous inequality is satisfied by a region of boxes which (initially) satisfy the Uffink inequality (see Fig. \eqref{fig:Uffink}). Thus, all boxes in this region can be distilled out of the Uffink set, implying that the latter is not closed.

Moreover, this implies that the set of correlations that violate the principle of information causality (IC) \cite{IC} can be extended to the shaded regions of Fig. \eqref{fig:Uffink}. This also implies that the set of correlations satisfying IC must be contained in the largest closed subset of the Uffink set.

\end{document}